# Quantum Proton Tunneling in Multi-electron/-proton Transfer Electrode Processes


*Ken Sakaushi* [a,b*]

[a] Center for Green Research on Energy and Environmental Materials, National Institute for Materials Science, 1-1 Namiki, 305-0044 Tsukuba, JAPAN

[b] Global Research Center for Environment and Energy Based on Nanomaterials Science, National Institute for Materials Science, 1-1 Namiki, 305-0044 Tsukuba, JAPAN

**Corresponding to KS**: sakaushi.ken@nims.go.jp


Quantum proton tunneling (QPT) in two representative multi-electron/-proton transfer electrode processes, *i.e.* hydrogen evolution reaction (HER) and oxygen reduction reaction (ORR), was investigated using polycrystalline platinum (pcPt) and gold (pcAu) electrodes at 298 kelvin (K). To observe quantum effects in the electrode processes, the hydrogen/deuterium kinetic isotope effect constant ratio ($\equiv K^{H/D}$) was measured in various conditions. For the HER in both acidic and alkaline conditions, results show that the pcPt exhibits a negligible or weak QPT evident by the small value of $K^{H/D}$ ($1 < K^{H/D} < 3$), which indicates that the semiclassical transition state theory (SC-TST) scheme dominates the rate-determining step (RDS). For pcAu in an alkaline condition, the $K^{H/D}$ was a small value of ca. 1 at a low $\eta$ region around 0.2 V. However, at a high $\eta$ region >



0.6 V, a high $K^{H/D}$ (> 13) was obtained. These results suggest a transition of the electrode process from SC-TST to a full QTP in the RDS on increasing the overpotential. For ORR with pcPt, $K^{H/D}$ higher than the theoretical maximum in SC-TST was observed in the alkaline condition at a low overpotential region. A primitive but robust theoretical analysis suggests that the QPT governs the rate-determining step of ORR in this condition. However, this full QPT path transits to the classical in a higher overpotential region. Therefore, contrary to the HER on pcAu in alkaline, the electrode process shows a transition from a full QPT to SC-TST on increasing the overpotential. No QPT in ORR on a pcPt electrode was observed in an acidic condition. This report describes that the QPT in surface electrochemical systems is strongly affected by the choice of system. Although several systems show a clear manifestation of QPT in the electrode processes and also primitive interpretations can be made of these observations, deriving a fine molecular-level picture of the results including several complicated effects remains challenging. However, the observations suggest that selection of a full QPT path might be affected strongly by different microscopic proton transfer mechanisms, *i.e.* proton transfer from hydronium ion or water molecules.



## 1. Introduction

Tunneling phenomena constitute a traditional topic in quantum physics.[1] They can be traced back to the Hund's seminal report in 1927,[2] which introduced an idea of barrier penetration in the framework of quantum mechanics to explain the theory of molecular spectra. During efforts to build a theoretical interpretation of alpha decay, quantum tunneling theory was developed by numerous physicists. Quantum tunneling is also known as a part of chemical and biological systems.[3-8] Furthermore, this effect has led to the invention of modern technologies for semiconductor devices and microscopy.[9, 10]

In 1931, R. W. Gurney[11] first applied quantum tunneling in electrode processes to explain electron transfer from an electrode to a proton in solution as a key step of kinetics in hydrogen evolution reactions. This view was modified in a much more realistic form by J. A. V. Butler,[12] *i.e.* electron transfer *via* radiationless quantum tunneling to a proton to form an adsorbed hydrogen (H*) on an electrode surface. Further developments were made by many other researchers. In addition to electron tunneling, the formulation of quantum proton tunneling (QPT) in electrode processes was presented in the work of Bawn and Ogden.[13] Later, detailed theoretical discussions were advanced by S. G. Christov and many other researchers.[14-16] These earlier theories emphasized initial activation processes because of anharmonic oscillations and the quantum distribution of



vibrator levels of nuclei, such as the protons within water or the hydroxonium ion. This view is so-called "thermal excitation model (or bond-stretching model)".[15] However, other theories to describe QPT in electrode processes are based on a continuum electrostatic approach put forth by Libby and Weiss,[17, 18] whose model was developed further to electrochemical formalisms based on the Marcus–Hush theory[19-22] or on the Dogonadze–Kuznetsov–Levich theory.[23-27] These pictures are classified as "medium(solvent)-reorganization model", which is essentially different from the thermal excitation model.[28] These two basic classes of proton transfer theories have inspired to develop different models such as the Anderson–Newns–Schmickler model[29-31], or have developed to advanced models, for instance the Soudackov–Hammes–Schiffer model.[32, 33] In addition, first-principle-based or path-integral molecular dynamics approaches[34-39] t or a grand canonical ensemble based approach for electrochemical reaction rate computation[40] to have been proposed to investigate QPT. In addition to theoretical contributions there are an enormous number of experimental efforts at observation and elucidation of QPT in electrode processes, as exemplified by works reported by electrochemistry experts such as Conway, Christov, or Krishtalik.[28, 41-48] Therefore, many previous efforts have been undertaken as both experimental and theoretical works to unveil quantum effects in electrode processes. Nevertheless, it remains unclear in which system we can observe QPT in electrode processes and why QPT emerges. Furthermore, the theoretical framework for the microscopic mechanism of QPT in electrode



processes is still under development. This is the major reason to underscore the challenge to elucidating the mechanisms of electrode processes in general, which has a quite long research history extending back to Faraday's era.[49] Therefore, in contrast to other technologies, modern quantum principles are not often applied for electrochemistry-oriented energy conversion technology as the core mechanism and to improve the properties.[50]

This report briefly presents an investigation of QPT in multi-electron/-proton transfer electrode processes with the present rudimentary understanding of the mechanism. As the model reactions, hydrogen evolution reaction (HER) and oxygen reduction reaction (ORR) were studied using polycrystalline platinum (pcPt) and gold (pcAu) electrodes at 298 kelvin (K) in acidic and alkaline conditions. It will be presented that the QPT are visible in several systems. Possible mechanisms will be discussed.

## 2. Present Understanding of Quantum Proton Tunneling in Electrode Processes

As described above, interest in quantum tunneling in electrochemistry began to appear in the early $20^{th}$ century during the establishment of quantum mechanics. From this long history of investigation, one can find a wide spectrum of detailed reports on electrode processes using H-D exchange isotope effects to date.[41, 43, 44, 48, 51-55] Furthermore, various theoretical models are



available. Nevertheless, QPT in electrode processes is still poorly understood because of several experimental challenges in electrochemical experiments. Unlike ordinary water systems,[56] this made it challenging to carry out precise electrochemical analyses using deuterated water, which is necessary to study quantum effects in electrode processes. For example, the first report to present a comparison of cyclic voltammograms in fully deuterated and protonated systems was of a study by Yeager and co-workers in 1985.[57] Some studies used tritiated water however it is far difficult to handle to use in highly clean electrochemical systems even compared to deuterated water because of its radioactive nature. Furthermore, as presented in Section 3, experimental procedures and analytical equations to observe and analyze QPT in electrode processes have only been established quite recently.[58, 59] In actuality, very little is known about QPT in molecular-level electrode processes. Therefore, we should begin experimental studies of this topic from an embryonic stage. Moreover, the theoretical description of microscopic activation processes on the electrode surface remains unclear. Indeed, this point is the objective of this contribution to share and discuss the protocol and analytical equations to observe and analyze QPT in electrode process, and to develop its present understanding by integration with modern theories and experimentally obtained results from related subjects.



## 3. Experimental Methods and Analytical Equations

## 3.1. Standard Electrochemical Methods

Electrochemical measurements were conducted using an RRDE set-up (Dynamic Electrode HR-301; Hokuto Denko Corp.) with an electrochemical analyzer (HZ-7000; Hokuto Denko Corp.) based on a custom-made three-compartment electrochemical glass cell at 298 K ± 1. The ring electrode was kept at 1.2 V *vs.* reversible hydrogen electrode (RHE) or reversible deuterium electrode (RDeE). We respectively denote potential with RHE scale and with RDeE scale as $V_{RHE}$ and $V_{RDeE}$. The cell was first cleaned by boiling in a mixture of concentrated sulfuric acid overnight and then boiling in ordinary ultrapure water (MilliQ water, 18.3 MΩ cm; Millipore Merck Corp.) overnight. Before electrochemical measurements, the cells were washed with an ultrapure electrolyte several times. Then they were filled with the ultrapure electrolyte for measurements. The electrolytes in the cell were bubbled respectively with $O_2$ (purity > 99.999%; Taiyo Nippon Sanso Corp.) or Ar (purity > 99.99995%; Taiyo Nippon Sanso Corp.) for 30 min before the experiments to prepare the $O_2$-saturated condition or Ar-saturated condition. Resistance of electrochemical systems was measured before each experiment using impedance measurements. This value was used to correct $iR$-drop. As described herein, all potential values are $iR$-corrected. For the preparation of ultrapure electrolytes with ordinary water, high-purity KOH (semiconductor grade, 99.99% trace metals basis; Sigma-Aldrich Corp.) or $H_2SO_4$ (96%,



Ultrapur.; Merck and Co. Ltd.) was mixed with ultrapure water (Milli-Q water, 18.3 MΩ cm;

Millipore, Merck Corp.). Electrolytes based on the deuterated oxide were prepared by mixing

potassium deuteroxide solution (40 wt. % KOD in $D_2O$; Cambridge Isotope Laboratories Inc.) or

sulfuric acid-$d_2$ solution 96–98 wt. % in $D_2O$ (99.5 at. % D, Sigma-Aldrich Corp.) with high-

purity deuterated oxide ("100%" distilled $D_2O$, Sigma-Aldrich Corp.) to obtain ultrapure 0.1M

KOD or 0.05M $D_2SO_4$ in $D_2O$. To prepare the ultrapure deuterated electrolytes, the as-received

"100%" $D_2O$ was purified based on the method explained earlier.[59] These highly clean electrolytes

are indispensable for the measurement of KIE.

The three-electrode setup consists of a carbon counter electrode, a RHE or a RDeE as the

reference electrode, and a 0.5 cm diameter commercial fixed polyAu or polyPt working electrode

(Hokuto Denko Corp.). Therefore, the geometrical surface area of the electrodes is 0.196 $cm^2$.

Typically, we used a scan rate of 50 mV $s^{-1}$ and a rotation rate of 1600 rpm for ORR experiments,

and a scan rate of 1 mV $s^{-1}$ and a rotation rate of 2000 rpm for HER/ deuterium evolution reaction

(DER). An electrochemically active surface area (ECSA) was measured using the typical method

to use cyclic voltammogram (CV) to normalize obtained currents for reliable comparison.[60] A

charge of 220 μC $cm^{-2}$ is assumed for a charge of full coverage of monolayer proton and deuteron

on a smooth polyPt surface. In the case of polyAu, a value of 424 μC $cm^{-2}$ is assumed as the



charge per unit area because of reduction of a monolayer of the surface oxide.

## 3.2. Standard Analytical Equations to Study Kinetics of Electrode Processes

The fundamental meanings of the key electrochemical parameters in this section are described in

the Appendix 1. In Section 3.4., the list of electrochemical parameters is available.

In this manuscript, we set two fundamental postulations: (1) there is Arrhenius' law in

overpotential ($\eta$) vs. common logarithm of kinetic current densities ($\log j_k$); and (2) there is only

a single rate-determining step for multistep electrode processes. Because of these postulations,

we yield following relation for multistep cathodic reactions, such as hydrogen evolution reaction

and oxygen reduction reaction

$$j_k = j_0 \exp\left(-\frac{\alpha F}{RT}\eta\right), \qquad\qquad\qquad \text{Eq. 1}$$

where $j_0$, $\alpha$, $F$, $R$, $T$, and $\eta$ respectively represent the exchange current density, the transfer

coefficient, the Faraday's constant, the gas constant, temperature, and overpotential. The

overpotentials is defined as $\eta$ = measuring potentials – the equilibrium potential of a reaction. The

exchange current density is defined as a current density at $\eta = 0$.



We define Tafel slope ($b$) in $\eta$-log $j_k$ relations, which is obtained by modifying Eq 1 to the Tafel's equation form

$$\eta = -b \, (\log \, j_k + \log \, j_0),$$

$$b \equiv \frac{2.303RT}{\alpha F}. \qquad\qquad \text{Eq. 2}$$

Furthermore, the transfer coefficient analysis based on Eq. 3 is applicable to acquire introductory information related to multistep electrode processes,[61] especially on rate-determining steps, where $s$, $v$, $\beta$, and $r$ respectively represent the transfer coefficient, the number of transferred electrons before the RDS, the stoichiometric number, the symmetric factor, and the number of transferred electrons in the RDS (usually $r = 1$).

$$\alpha = \frac{s}{v} + \beta r. \qquad\qquad \text{Eq. 3}$$

To use Eq. 3, as we mentioned already, we postulated that only a single RDS exists in each system. In fact, Eq. 3 assumes Tafel's law,[62] the absence of double-layer effects, and the low coverage of reactants/products on an electrode.



Coverage of proton or oxygen species on a surface is well known to affect the electrode kinetics. These terms are usually potential-dependent. In fact, they can be described as an exponential term together with $\beta$ for rate expression, for instance, as

$$-\frac{1}{b} = \frac{F}{RT}(\beta + \frac{RT}{F}\frac{\partial(\ln\theta_H)}{\partial\eta}),$$

for HER,[63] where $b$ and $\theta_H$ respectively denote the Tafel slope and the proton coverage on the electrode surface. From this, we give the third postulation to apply to Eq. 3: major microscopic effects attributable to the adsorbed species and electrified surface, which affect electrode processes, can be incorporated in $\beta$:

$$\beta = \beta' + \sum_{x=1} O(\eta^x).$$

Term $\beta'$ represents a symmetry factor in the traditional definition, which is often assumed as 0.5. However, this value can be varied if one considers the microscopic effects such as anharmonicities of the H* bond vibration.[64] The additional multicomponent term $\Sigma O(\eta^x)$ represents complex interactions of adsorbed species, electrified surface, solvent, etc., which can be varied with a



selection of theoretical framework. Symmetry factor $\beta$ has a central role in electrode kinetics. It is a fundamentally necessary entity to elucidate the microscopic mechanisms in both experimental and theoretical approaches. Therefore, because of the third postulation, it is challenging to obtain a detailed view on the microscopic mechanism. However, although it is a quite primitive approach and therefore we cannot directly identify major microscopic effects to emerge QPT, in this study, we specifically examine the observation of QPT and extract fingerprints of the main effects arising from quantum phenomena. This approach is expected to elucidate which theoretical frameworks can be applied to QPT in electrode processes, and then we would understand main reasons for quantum effects in a specific electrode process under these frameworks. Aiming to this objective, the simple framework described above is sufficient to pursue this study.

The ORR kinetic currents on Pt electrode is separable from diffusion limiting current using the following equation because we can observe clear diffusion-limited currents as

$$j_k = \frac{(j_{lim} \cdot j)}{(j_{lim} - j)},$$

where $j$ and $j_{lim}$ respectively represent the experimentally obtained current with the RRDE technique and a diffusion-limiting current.



The equilibrium potential for the $D_2O$ formation ($E^0_{D2O}$) is 1.262 $V_{RDeE}$.[57] Therefore, the overpotential for the ORR in a deuterated system ($\eta^D_{ORR}$) is obtainable by following the definition of overpotential:

$\eta^D_{ORR}$ (V) = an experimentally obtained potential ($V_{RDeE}$) – 1.262 ($V_{RDeE}$).

The overpotential for the ORR in the ordinary system ($\eta^H_{ORR}$) is obtainable as

$\eta^H_{ORR}$ (V) = an experimentally obtained potential ($V_{RHE}$) – 1.229 ($V_{RHE}$).

## 3.3. Analytical Equations to Obtain Kinetic Isotope Effect Rate Constant Ratio

To investigate the kinetic isotope effect (KIE) for HER, a KIE rate constant ratio in HER ($k^H(\eta)$ /$k^D(\eta) \equiv K^{H/D}_{HER}(\eta)$) is obtainable using

$$K^{H/D}_{HER}(\eta) = \frac{k^H(\eta)}{k^D(\eta)} = \frac{j_0^H}{j_0^D} \times \frac{[D^+]}{[H^+]} \times \exp\left\{\frac{(\alpha^D - \alpha^H)F\eta}{RT}\right\}, \qquad \text{Eq. 4}$$

where $k_0$, $[H^+]$, $[D^+]$, and $T$ respectively represent a rate constant at $\eta = 0$, proton concentration,



deuteron concentration, and an experimental temperature (298±1 K in this study). Detailed derivation of the equation is presented in the Appendix 2. Herein, we note that we must incorporate consideration of the different dissociation constant of $D_2O$ from that of $H_2O$.[54] This simple difference has been underestimated in several studies.[52, 53, 65] Nevertheless, it is an extremely important quantity to analyze KIE in electrode processes, as shown in the following equations. For example, the 0.1M KOD in $D_2O$ solution gives $[D^+] = 10^{-13.87}$ mol $L^{-1}$. Therefore, the $K^{H/D}_{HER}(\eta)$ for the case comparing the 0.1M KOH in $H_2O$ ($[H^+] = 10^{-13}$ mol $L^{-1}$) and 0.1M KOD in $D_2O$ solutions is

$$K^{H/D}_{HER}(\eta) = 0.1349 \times \frac{j_0^H}{j_0^D} \times \exp\left\{\frac{(\alpha^D - \alpha^H)F\eta}{RT}\right\}.$$

However, the $K^{H/D}_{HER}(\eta)$ for the case comparing 0.05M $H_2SO_4$ in $H_2O$ ($[H^+] = 10^{-1}$ mol $L^{-1}$) and 0.05M $D_2SO_4$ in $D_2O$ ($[D^+] = 10^{-1}$ mol $L^{-1}$) becomes

$$K^{H/D}_{HER}(\eta) = \frac{j_0^H}{j_0^D} \times \exp\left\{\frac{(\alpha^D - \alpha^H)F\eta}{RT}\right\},$$

which is identical to the equation shown by Krishtalik as the general equation to obtain $K^{H/D}_{HER}(\eta)$.[65]



A KIE rate constant ratio for ORR ($K^{\text{H/D}}{}_{\text{ORR}}(\eta)$) is obtainable as

$$K^{\text{H/D}}{}_{\text{ORR}}(\eta) = \frac{k^{\text{H}}(\eta)}{k^{\text{D}}(\eta)} = \frac{j_0^{\text{H}}}{j_0^{\text{D}}} \times \frac{C_0^{\text{D}}}{C_0^{\text{H}}} \times \exp\left\{\frac{(\alpha^{\text{D}} - \alpha^{\text{H}})F\eta}{RT}\right\},$$  Eq. 5

where $C_0$ denotes the oxygen concentration. Also, $C_0^{\text{D}}/C_0^{\text{H}}$ is known to be 1.101 at 298 K.[52]

### 3.4. List of Parameters

Meanings of key electrochemical parameters are available in Appendix 1.

Transfer coefficient: $\alpha$, Symmetry factor: $\beta$, Stoichiometric number: $v$, Overpotential: $\eta$, The number of transferred electrons before the RDS: $s$, The number of transferred electrons in the RDS: $r$, Kinetic current density: $j_k$, Exchange current density: $j_0$, Limiting current density: $j_{\text{lim}}$.

## 4. Results and discussion

### 4.1. Quantum Proton Tunneling in Hydrogen Evolution Electrode Process

First, the HER activities and electrode processes of pcPt in 0.05M $H_2SO_4$ in $H_2O$/0.05M $D_2SO_4$ in $D_2O$ are discussed based on the diagrams of an overpotential ($\eta$) vs. common logarithm of the



*iR*-corrected kinetic HER current densities (log $j_k$) (Figure 1a). We used current densities normalized by the electrochemically active surface areas (ECSA) of the electrocatalysts.[60] To discuss the mechanism of HER in acid, we follow two well-accepted paths:[63, 66] the primary discharge step (or so-called Volmer step, where * denotes adsorbed species or adsorption sites) is

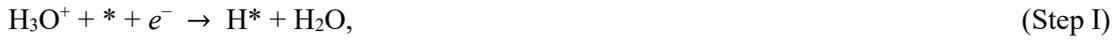

(Step I)

coupled with either the electrochemical-desorption step (Heyrovský step)

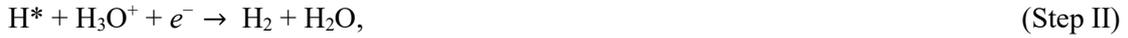

(Step II)

or the recombination step (Tafel step)

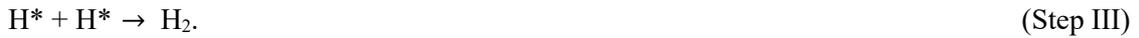

(Step III)

The results showed that the Tafel slopes (*b*) are slightly different for the ordinary and deuterated systems, respectively showing $-27.4 \pm 0.2$ mV/dec and $-29.9 \pm 0.4$ mV/dec, which are in good agreement with earlier reports describing that *b* = ca. 30 mV/dec for HER on Pt in acidic conditions (Figure 1a).[67, 68] A small difference of *b* in the deuterated and ordinary systems might be explained by the difference of H* and D* bond vibrations, which engenders different anharmonicities and therefore different symmetry factors, or other microscopic effects. From Eq. 2, we obtain transfer coefficients (*α*) = ca. 2 for both ordinary and deuterated systems. With *α* and Eq. 3, one can obtain additional details on the rate-determining step (RDS) and KIE. When *α* = 2,



the possible combination of $(s, v, r)$ in Eq. 3 is $(2, 1, 0)$ because of $0 < \beta < 1$ and typically $\beta = 0.5$, which indicates that the possible RDS of HER on pcPt in 0.05M $H_2SO_4$ in $H_2O$/0.05 M $D_2SO_4$ in $D_2O$ is Step III (the recombination of adsorbed protons (H*) /deuteron (D*) on the electrode surface), which shows good agreement with numerous earlier reports describing the HER activity of Pt in acidic conditions using ordinary water.[63]

For the analysis of KIE, the pcPt electrode in the acidic condition shows $K^{H/D}_{HER}(\eta)$ of about 2 in the overpotential range of $-0.005$ to $-0.02$ V, which corresponds to the linear region in 0.05M $H_2SO_4$ solution (Figure 1a). The criteria to identify a QPT path or a SC-TST path in this report are based on earlier discussions of theoretical situations for proton transfer with or without tunneling.[69, 70] In the framework of Westheimer and Melander,[69] the maximum $K^{H/D}$ without QPT is about 10. On the other hand, the Kiefer–Hynes (K-H) formalism suggests that the maximum $K^{H/D}$ without QPT is about 6 at 300 K.[70] Therefore, in the case of $K^{H/D} > 10$, we conclude that it is a full QPT path. As discussed already, $b$ shows slightly different values in the deuterated and ordinary water systems. Therefore, this engenders $\alpha^H \neq \alpha^D$ for Eq. 4. In this case, a KIE in electrochemistry is readily apparent as a function of potential. This result suggests that a small difference between $\alpha^H$ and $\alpha^D$ can strongly affect KIE. As shown in Figure 1b, $K^{H/D}_{HER}$ is 1.74 at $\eta = -0.005$ V and increases to 1.93 at $\eta = -0.02$ V, suggesting an 11% increase of $K^{H/D}_{HER}$ with



0.015 V of an overpotential shift. However, although $K^{H/D}_{HER}$ increases with the shift of overpotential, the small value of $K^{H/D}_{HER}$ is explainable with the semiclassical transition-state theory (SC-TST) or no proton transfer in the RDS. Therefore, no QPT exists in this condition.

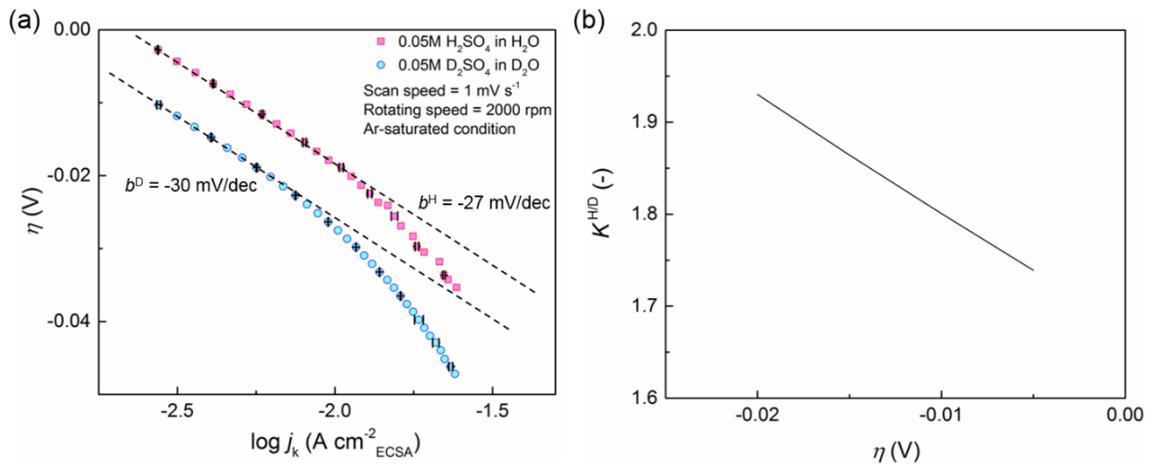

**Figure 1.** (a) $\eta$ vs. log $j_k$ diagrams and (b) $K^{H/D}_{HER}$ vs. $\eta$ diagram of the pcPt in 0.05 M $H_2SO_4$ in $H_2O$/0.05 M $D_2SO_4$ in $D_2O$. Kinetic current densities ($j_k$) are normalized by ECSA (A cm$^{-2}_{ECSA}$). All data are corrected for ohmic drops ($iR$ correction). Measurements were performed with a scan rate of 1 mV s$^{-1}$ and a rotation rate of 2000 rpm. Error bars are shown for every third data point.

Next, the HER activities and electrode processes of pcPt in 0.1M KOH in $H_2O$/0.1M KOD in $D_2O$ are discussed based on the $\eta$ vs. log $j_k$ (Figure 2a). In alkaline conditions, water molecules become the proton source.



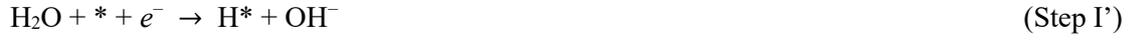
$$\text{H}_2\text{O} + \ast + e^- \ \rightarrow \ \text{H}\ast + \text{OH}^- \hspace{4cm} \text{(Step I')}$$

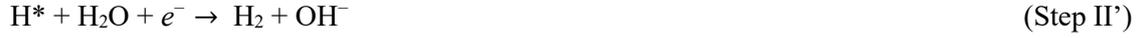
$$\text{H}\ast + \text{H}_2\text{O} + e^- \ \rightarrow \ \text{H}_2 + \text{OH}^- \hspace{3.5cm} \text{(Step II')}$$

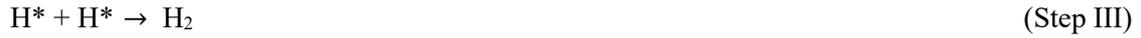
$$\text{H}\ast + \text{H}\ast \ \rightarrow \ \text{H}_2 \hspace{5cm} \text{(Step III)}$$

A clear line exists in both the ordinary and deuterated systems (Figure 2a). The Tafel slopes ($b$) are identical, respectively showing $-69.2 \pm 0.1$ mV/dec and $-69.1 \pm 0.2$ mV/dec. This result also agrees well with earlier reports that have described $b$ in alkaline conditions of between 50 and 75 mV/dec.[71, 72] From Eq. 2, we obtain the transfer coefficient ($\alpha$) = 0.86. When $\alpha$ = 0.86, the possible combinations of ($s$, $v$, $r$) in Eq. 3 are (0, 2, 1) or (1, 1, 1), which gives $\alpha = \beta$ and $\alpha = 1 + \beta$, respectively. We postulated that only a single RDS exists in the total process (section 3.2). In the case of $\alpha = \beta$, the limiting step is Step I', with $\beta$ = 0.86. In contrast, in the case of $\alpha = 1 + \beta$, the limiting step is Step II', with $\beta$ = 0.36. From earlier reports,[63, 64] we can expect that Step I' is quasi-equilibrium. Therefore, Step II' can be the possible RDS.

To examine details of the electrode process in this system further, KIE was checked. The KIE of the pcPt electrode in the alkaline condition shows a hydrogen/deuterium kinetic isotopic rate constant ratio ($K^{\text{H/D}}_{\text{HER}}(\eta)$) close to 1. The $b$ shows the identical value of $-69$ mV/dec in the



deuterated and ordinary systems. Therefore, we obtained potential independent $K^{H/D}_{HER}$ because of $\alpha^H = \alpha^D$ for Eq. 4 (Figure 2b). In the case of $K^{H/D} = 1$, it is the sign for no involvement of proton transfer in the RDS. Therefore, combined with the discussion presented above based on Eq. 2, the RDS in this system can be an electron transfer process related to Step II'. Nevertheless, no QPT exists in this condition. Therefore, we do not discuss the detailed mechanisms of this condition further.

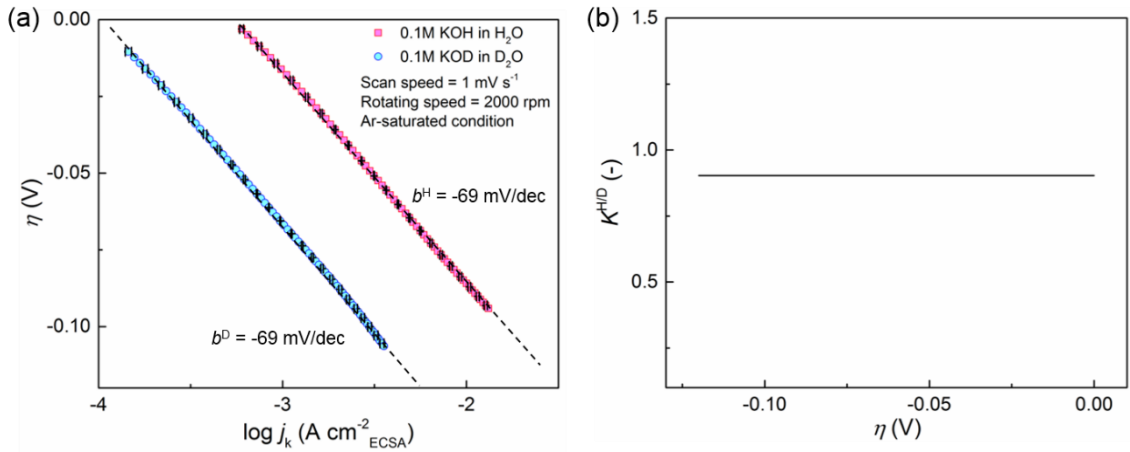

**Figure 2.** (a) $\eta$ vs. $\log j_k$ diagrams and (b) $K^{H/D}_{HER}$ vs. $\eta$ diagram of the pcPt in 0.1M KOH in $D_2O$/0.1M KOD in $D_2O$. Kinetic current densities ($j_k$) are normalized by ECSA (A cm$^{-2}_{ECSA}$). All data are corrected for ohmic drops ($iR$ correction). Measurements were taken with a scan rate of 1 mV s$^{-1}$ and a rotation rate of 2000 rpm. Error bars are shown for every third data point.

Next, we move to the system with a pcAu electrode. In Figure 3, the HER activities and electrode



processes of pcAu in 0.05M $H_2SO_4$ in $H_2O$/0.05M $D_2SO_4$ in $D_2O$ are shown. In this condition, a clear line exists in both systems. The Tafel slopes ($b$) are identical, showing $-82.2 \pm 0.7$ mV/dec and $-81.8 \pm 0.2$ mV/dec, respectively, in 0.05M $H_2SO_4$ in $H_2O$ and 0.05M $D_2SO_4$ in $D_2O$. From Eq. 2, we obtain the transfer coefficients ($\alpha$) = 0.72, which give the possible combinations of ($s$, $v$, $r$) in Eq. 3 as (0, 2, 1) or (1, 1, 1) leading respectively to $\alpha = \beta$ and $\alpha = 1 + \beta$. In contrast to the Pt electrode, we observe negligible UPD-/OPD-H in the case of an Au electrode. Therefore, the quasi-equilibrium for Step I is not a reasonable assumption. The case of ($s$, $v$, $r$) = (1, 1, 1), *i.e.* Step II as the RDS, is unlikely for this system, at least in a low $\eta$ region. To consider the RDS in this system, we examine earlier reports,[73] which used a first-principle-based calculation to consider the proton adsorption and HER on metal electrodes. These reports suggest that proton adsorption on the Au(111) surface is endergonic at the equilibrium potential (*i.e.* $\eta = 0$) for the HER. This result indicates that Step I determines the rate of the overall reaction as the potential-determining step.[54, 74] The $K^{H/D}_{HER}(\eta)$ of this system is calculable to $3.0 \pm 0.3$ with no potential-dependency because of $\alpha^H = \alpha^D$. Although detailed information related to H* bond length and the activation barrier height and width at the RDS are required even for a primitive estimation of a tunneling magnitude for an overall reaction, $K^{H/D} = 3$ suggests that the QPT can affect the electrode kinetics, although it might be a minor effect. It is noteworthy that it is not straightforward to distinguish QPT and nontunneling proton transfer.[70, 75, 76]



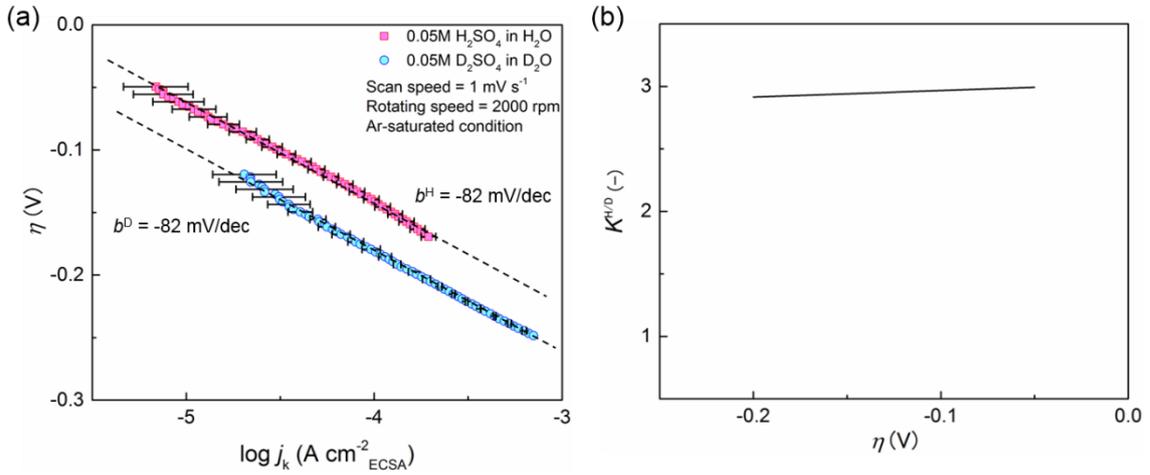

**Figure 3.** (a) $\eta$ vs. $\log j_k$ diagrams and (b) $K^{H/D}_{HER}$ vs. $\eta$ diagram of the pcAu in 0.05M $H_2SO_4$ in $D_2O$/0.05M $D_2SO_4$ in $D_2O$. Kinetic current densities ($j_k$) are normalized by ECSA (A cm$^{-2}_{ECSA}$). All data are corrected for ohmic drops ($iR$ correction). Measurements were taken with a scan rate of 1 mV s$^{-1}$ and a rotation rate of 2000 rpm. Error bars are shown for every third data point.

Finally, in Figure 4, the HER activities and the electrode processes of the pcAu electrode in 0.1M KOH in $H_2O$/0.1M KOD in $D_2O$ are shown. In this condition, we can observe a clear line in each system for widely various overpotentials. The values of the Tafel slopes differ greatly in the two systems. In fully protonated and the fully deuterated systems, the Tafel slopes ($b$) are, respectively, $101.7 \pm 0.2$ mV/dec and $135.2 \pm 0.6$ mV/dec (Figure 4a). From Eq. 2, we obtain the transfer



coefficients in the ordinary hydrogen system and the fully deuterated system respectively as $\alpha^H = 0.58$ and $\alpha^D = 0.44$. Based on these results, we assigned $(s, v, r)$ in Eq. 3 as (0, 0, 1), and obtain $\alpha = \beta$. Therefore the RDS is Step I'. The $K^{H/D}_{HER}(\eta)$ in this system shows an anomalous behavior. The $K^{H/D}_{HER}(\eta)$ is close to 1 at $\eta = -0.2$ V, suggesting that no contribution of proton transfer or proton transfer with SC-TST occurs in the RDS. However, $K^{H/D}_{HER}(\eta)$ increases drastically to > 13 at $\eta = -0.7$ V, which might be interpreted as a sign of the full quantum process (Figure 4b). Therefore, in this system, the electrode process can show a transition from a (semi)classical process to the quantum process associated with an increase of overpotential.

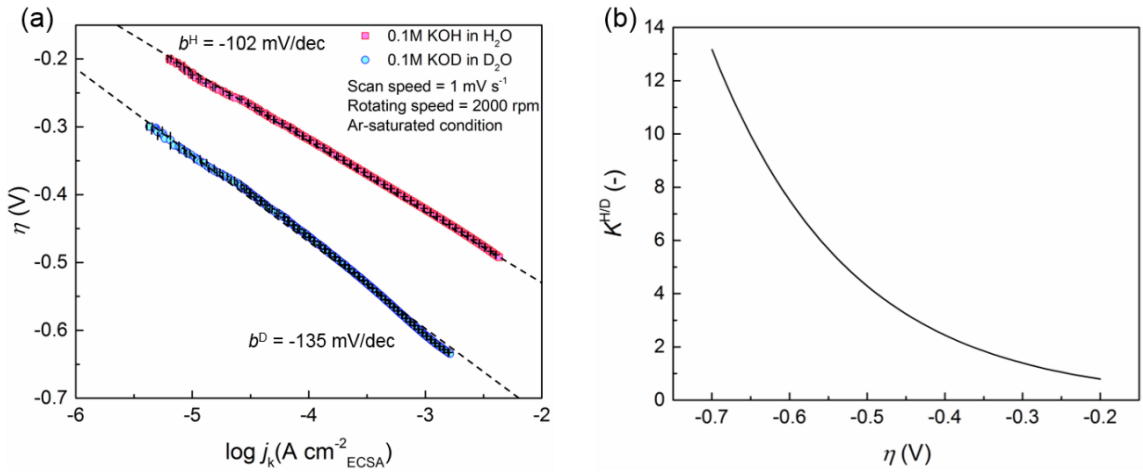

**Figure 4.** (a) $\eta$ vs. log $j_k$ diagrams and (b) $K^{H/D}_{HER}$ vs. $\eta$ diagram of the pcAu in 0.1M KOH in D$_2$O/0.1M KOD in D$_2$O. Kinetic current densities ($j_k$) are normalized by ECSA (A cm$^{-2}_{ECSA}$). All data are corrected for ohmic drops ($iR$ correction). Measurements were taken with a scan rate of 1 mV s$^{-1}$ and a rotation rate of 2000 rpm. Error bars are shown for every third data point.



**4.2. Summary of Quantum Proton Tunneling in Hydrogen Evolution Electrode Process**

The key results in hydrogen evolution electrode processes are presented in Table 1. The results show that, beside the pcPt electrode in the alkaline condition, the $K^{H/D}$ values suggest there are fingerprints of possible quantum effects in proton transfer in the other three systems. Especially, the most interesting system among these three is the pcAu electrode in the alkaline condition because of the observation of classical-to-quantum transition of proton transfer on increasing overpotential: In this system, proton transfer is negligible initially because $K^{H/D}$ is close to 1 at $\eta$ = −0.2 V. However, its value increases drastically to > 13 at $\eta$ = −0.7 V, suggesting domination of QPT, *i.e.* a full QPT path. We note that the maximum $K^{H/D}$ at 298 K without tunneling is predicted from 10 to 13 depended on a selection of models,[69, 70, 75, 77] therefore $K^{H/D}$ > 10 regarded as a manifestation of a full QPT path.



**Table 1.** Key kinetic values and analysis for HER on pcPt and pcAu

| Electrode | Condition | $\alpha$ | RDS | $-\log j_0$ (A cm$^{-2}_{\text{ECSA}}$) | $K^{\text{H/D}}$ |
|---|---|---|---|---|---|
| pcPt | 0.05M H$_2$SO$_4$ in H$_2$O | 2 | Step III | 2.66 | ca. 1.8 |
|  | 0.05M D$_2$SO$_4$ in D$_2$O | 2 | Step III | 2.88 | $\eta$ dependent negligible QPT |
|  | 0.1M KOH in H$_2$O | 0.86 | Step II'* | 3.24 | 1 |
|  | 0.1M KOD in D$_2$O | 0.86 | Step II'* | 4.07 | $\eta$ independent no QPT |
| pcAu | 0.05M H$_2$SO$_4$ in H$_2$O | 0.72 | Step I | 5.73 | 3 |
|  | 0.05M D$_2$SO$_4$ in D$_2$O | 0.72 | Step I | 6.21 | $\eta$ independent weak QPT |
|  | 0.1M KOH in H$_2$O | 0.58 | Step I' | 7.15 | 1 to > 13 |
|  | 0.1M KOD in D$_2$O | 0.44 | Step I' | 7.55 | $\eta$ dependent classical-to-quantum transition |

*Because of $K^{\text{H/D}} = 1$, the RDS is the electron transfer step in Step II'.

A key factor for a full QPT path in the system is the large difference of $\alpha^{\text{H}}$ and $\alpha^{\text{D}}$ as shown in Eq. 4. Because data related to H* bond and the activation barrier for Step I' in the corresponding system remain insufficient, it is difficult to conduct a quantitative analysis. However, qualitatively, the difference of $\alpha$ ($= \frac{s}{\nu} + \beta$) in the H and D systems suggests that the harmonic approximation and/or ignorance of microscopic interactions (*i.e.* the multicomponent interaction term $\Sigma O(\eta^{\text{x}})$ in section 3.2) would not be suitable approaches to consider electrochemical systems found to have significant quantum effects on proton transfer. In the classical framework, we assume that the



symmetry factor is $\beta = 0.5$ because of the harmonic approximation and no microscopic interactions (*i.e.* $\Sigma O(\eta^x) = 0$). However, a wide spectrum of interactions exists between electrified surface effects, adsorbed species, and solvents. Therefore, the present results indicate that the theoretical framework to consider the QPT in electrode process should be constructed with careful selection of interactions and an addition of anharmonicities in the reaction scheme (Figure 5). Recent results of surface enhanced infrared spectrometry (SEIRAS) on Au electrode surfaces present a primitive picture at a microscopic level of Step I'.[78-81] The applied potential below the potential of zero charge (PZC) of an Au electrode surface suggests that the orientation of water molecules can be inferred as organized with the two hydrogen atoms slightly closer to the electrode surface than the oxygen atom (Figure 5a). Furthermore, we should incorporate effects of the surface reconstruction in the case of Au electrodes.[82, 83] As demonstrated in previous works, the first monolayer of an Au surface will be developed as associated with potential change.[83-85] For example, in the case of an application of negative potential to the most thermodynamically stable Au (111) surface, the Au(111)-(1×1) phase transitions to the Au(111)-(23×√3) uniaxial striped phase. This fact indicates that the surface phase of Au electrode will be developed on changing the electrode potential. Recent reports suggest that a lattice strain can have a strong effect on electrocatalytic activity.[86, 87] Therefore not only anharmonicities and microscopic interactions but also surface structure evolution might be important to account for. Of course,



these different effects mutually interact to change the activation barrier on applying potential (Figures 5b and 5c). Last but not least, a recent report of a study using an Au electrode describes a large difference of the transfer coefficients between protonated and deuterated systems.[88] Reportedly, the RDS corresponding to this system is a proton-coupled electron transfer. Their experimentally obtained results were well fitted with a model considering contributions from excited electron–proton vibrionic states,[89] which are affected by isotopes and by changing potential. Although HER is the simplest multi-electron/-proton transfer electrode process, the system complexity increases enormously if one tries to ascertain the quantum effects in its microscopic mechanism.



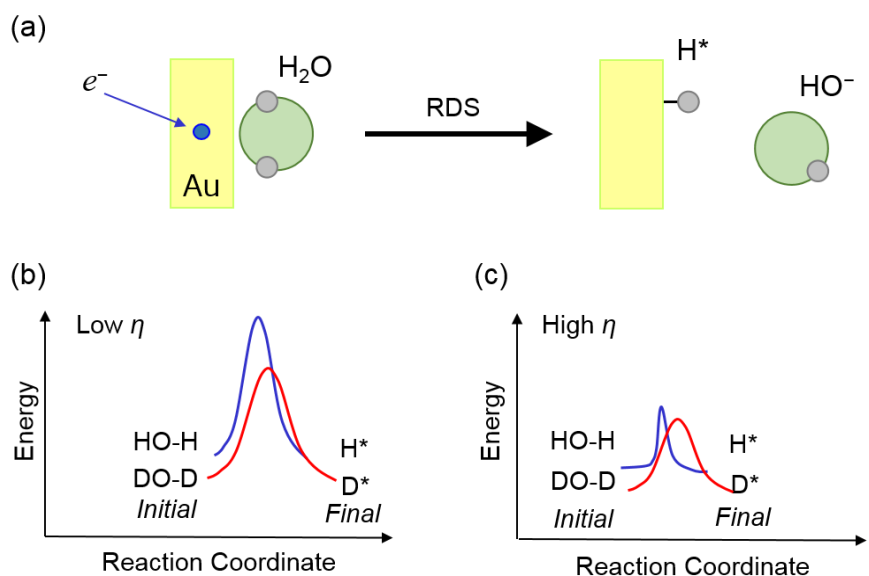

**Figure 5.** (a) Schematic illustration for the Step I' ($H_2O + e^- + * \rightarrow H* + OH^-$) on an Au electrode with a possible microscopic mechanism. Water molecule in an alkaline condition is postulated to be adsorbed onto an Au electrode with the same configuration as that of an acidic condition obtained by SEIRAS.[79, 81] The proton source of HER in alkaline condition is water molecule. Therefore, the adsorbed water molecule is assumed to contribute to Step I'. Hypothetical reaction coordinates (b) at low $\eta$ and (c) at high $\eta$ are based on the KIE results. The different peak positions in the protonated and deuterated systems are attributable to different symmetry factors ($\beta^H = 0.58$ and $\beta^D = 0.44$). The Brønsted–Evans–Polanyi relation with these symmetry factors predicts that the barrier height in the protonated system should be decreased greatly from low to high $\eta$ compared to that of the deuterated system. Therefore, to produce a full QPT path and a large $K^{H/D}$ at a high $\eta$, the barrier width in the protonated system is predicted to become sufficiently thin on increasing $\eta$ to increase the tunneling probability.



**4.3. Quantum Proton Tunneling in Oxygen Reduction Electrode Process**

To date, QPT in ORR was observed in the case of pcPt in an alkaline condition (Figure 6).[58] In this case, QPT was confirmed in a low overpotential region because of $K^{H/D} = 32$, but the QPT transitions to a SC-TST scheme ($K^{H/D} = 3.7$). Therefore, a quantum-to-classical transition of proton transfer in ORR (Figure 6a) was confirmed using theoretical analysis (Figure 6b). The QPT fingerprint was indicated by a large $K^{H/D} > 13$ in the low overpotential region: The maximum $K^{H/D}$ for the O–H bond breaking in semiclassical frameworks is ca. 13 at 298 K, accounting for the change in the activation barriers because of differences in zero-point energies attributable to different vibrations in O–H and O–D.[77] In addition to this, it is noteworthy that, although different adsorption energies of OH/OD species on Pt surface can cause differences in zero-point energies or another reaction path,[90, 91] the difference of the OH/OD adsorption energies in the system is 1.2 kJ mol$^{-1}$, which is too small to affect the reaction. In this study, a primitive but robust model was applied to analyze this observation by approximating the barrier by an asymmetric Eckart's one-dimensional potential energy function.[92, 93] Furthermore, the Brønsted–Evans–Polanyi relation was used to describe linear variations in the height of the activation barrier with the reaction energy on potential changing.[94-96] Although the barrier height for the corresponding reaction is under debate to the present day, the values from several report showing around 0.7–0.8 eV well agree with our observation.[97, 98] Furthermore, the barrier width was selected as 0.3 Å because the



widths, corresponding to hydrogen bonds between intermediate species such $O_2$* and O* are 0.25–0.35 Å. With this quite primitive model, QPT can be shown as possible. Also, a quantum-to-classical transition exists when increasing the overpotential (Figures 6b and 6c).



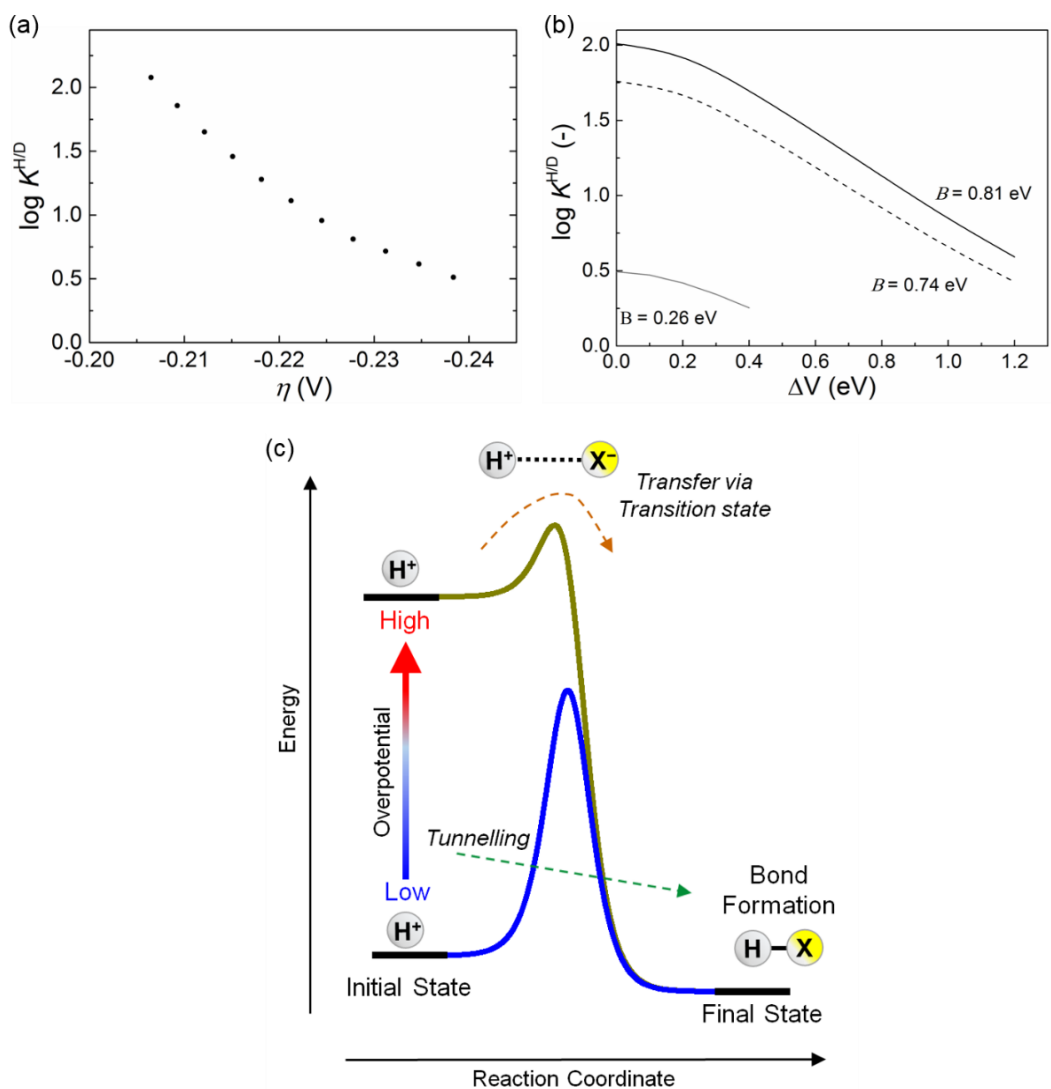

**Figure 6.** (a) $\eta$ vs. log $K^{H/D}_{ORR}$ diagram of the pcPt in 0.1M KOH in $D_2O$/0.1M KOD in $D_2O$, and

(b) theoretical log $K^{H/D}_{ORR}$ vs. reaction exothermicity parameter ($\Delta V$) diagram. The proton transfer

barriers at equilibrium ($B$) were referred from earlier reports. (c) Schematic diagrams for two

possible schemes of the proton-transfer reaction: proton transfer via SC-TST and QPT. The relative

contributions of the two mechanisms can be tuned on changing electrode potential. Figures reprinted

with permission from [*Phys. Rev. Lett.*, 121, 236001, (2018)] Copyright (2019) by the American

Physical Society.



In the case of pcPt in an acidic condition (Figure 7), no QPT was observed. Our experimentally

obtained result for $K^{H/D}$ was almost identical to the result obtained by Yeager and his co-workers,[57]

who conducted the same measurement and analysis for pcPt in a different acid (85% $H_3PO_4$). In

our case, the results showed that $K^{H/D} = 1.5$, with $K^{H/D} = 1.4$ found for Yeager's case, from which

it was concluded that the contribution of proton transfer is negligible: therefore the RDS is the

electron transfer process. This earlier report presented that conclusion because they conducted

their experiments for the protonated system at 298 K, but for the deuterated system at 303 K. This

report claims that $K^{H/D} = 1.4$ is the result of the 5 K difference, which engenders an increase in

the rate constant in the deuterated system by about 30% compared to the protonated system.

Furthermore, the K-H theory supports Yeager's conclusion because this theory predicts that the

minimum $K^{H/D}$ is about 2.[70] It should be emphasized that Yeager's study measured ORR in 85%

$H_3PO_4/D_3PO_4$ systems. It is therefore a different acidic system from that examined in the present

study. Also, they did not incorporate consideration of the different solubility of $O_2$ in the

protonated/deuterated systems at different temperatures. From these points obtained from a huge

number of experimental ORR studies, it is believed that the RDS of ORR on a Pt surface in an

acidic condition is related to the first electron transfer to dioxygen without association of proton

transfer,[99, 100] which leads to $K^{H/D} = 1$–2 owing to the K-H picture. However, from several

experimental studies, the RDS of ORR on Pt in acidic condition is the first proton-coupled



electron transfer, *i.e.* a proton transfer is included in the RDS.[101-104] Therefore, results showing

$K^{H/D}$ = ca. 1.5 can makes it difficult to conclude what is the RDS. Although continuous ongoing

efforts are made for understanding the microscopic mechanism of ORR, there are always updates

in the field.[105-117] Demands persist for both experimental and computational approaches to clarify

the microscopic electrode process of this reaction.[118, 119]

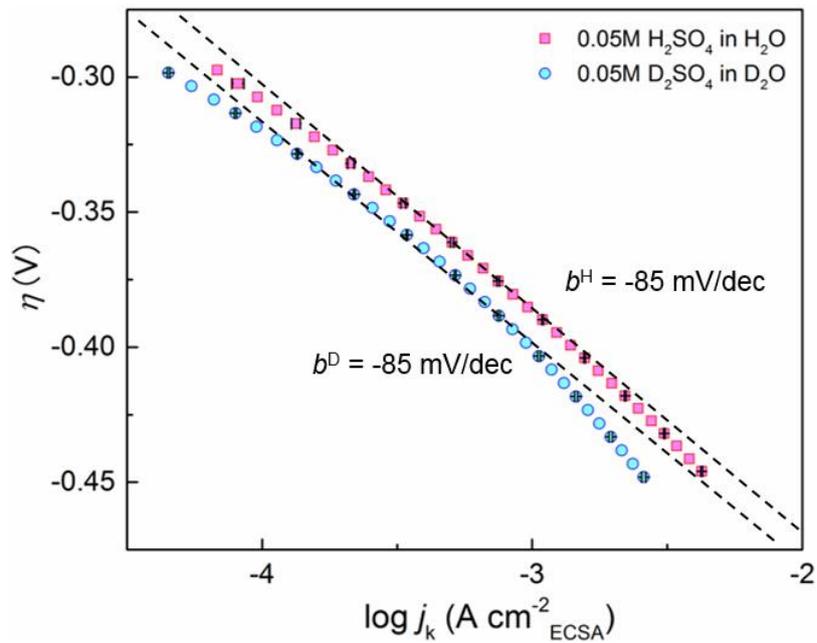

**Figure 7.** $\eta$ *vs.* log $j_k$ diagrams of the pcPt in 0.05M H$_2$SO$_4$ in D$_2$O/0.05M D$_2$SO$_4$ in D$_2$O. Kinetic

current densities ($j_k$) are normalized by ECSA (A cm$^{-2}_{ECSA}$). All data are corrected for ohmic drops

($iR$ correction). Measurements were taken with a scan rate of 50 mV s$^{-1}$ and a rotation rate of 2000

rpm. Error bars are shown for every third data point.



## 5. Conclusions

To observe QPT in multi-electron/-proton transfer electrode processes and to give fundamental interpretations of these observations at a microscopic level, we have established basic analytical equations and have prepared detailed experimental procedures to investigate quantum effects in electrode processes with the aid of kinetic isotope effects. Based on these approaches, we studied QPT in HER and ORR as model processes. The results have led to the following main conclusions:

(1) With the appropriate experimental procedures and equations, we can observe and analyze quantum effects in electrode processes. Results from this study indicate that QPT in electrode processes is not a rare phenomenon. This quantum process can be combined often with a SC-TST path. However, a full QPT process is apparently non-trivial.

(2) In our study, a full QPT in HER can be observed in only one condition: pcAu electrode in the alkaline condition at a high overpotential. Actually, QPT can be observed under other conditions, but they might be combined with the TST path. Under some conditions, no QPT was observed.

(3) The strong quantum effect in HER with pcAu electrode in the alkaline condition is attributable



to a huge difference of the symmetry factors between fully deuterated and protonated systems. The RDS is inferred as the first proton-coupled electron transfer for both the systems. The reason for the different symmetry factors might be a sum of several microscopic effects at the electrode/electrolyte interface, such as anharmonicities in the potential energy surface and/or changing adsorption energies of reaction intermediates because of surface reconstruction.

(4) The ORR using pcPt electrode in the alkaline condition shows a full QPT in a low overpotential region but no QPT in an acidic condition, at least in the overpotential range in this report.

(5) Many different forms of QPT are involved in electrode processes, such as the classical (SC-TST) to a full quantum scheme associated with an increase of overpotential (the HER on pcAu in the 0.1M KOH condition), and a full quantum to the classical scheme (the ORR on pcPt in the 0.1M KOH condition). At this stage, it is a challenge to give molecular-level interpretations to these QPT electrode processes because of the participation of many complicated effects. However, the different microscopic proton transfer mechanisms, *i.e.* proton transfer from hydronium ion or water molecule, might be a key to elucidating the full QPT path.



Overall, the experimental methods and equations presented herein enable QPT to be observed and analyzed in two model electrode processes, and provide primitive microscopic views of its mechanisms on electrolyte/electrode interfaces associated with quantum effects. To date, it is difficult to give fine molecular-level interpretations to observed results because it is unclear which effects are effectively contributing to the emergence of QPT in electrode processes. To comprehend the full microscopic picture of these complicated systems, it is necessary to continue integrating experimental and theoretical works to clarify these systems using state-of-the-art knowledge from physics and chemistry.

**Conflicts of interests**

The author has no conflict to declare.


**Acknowledgments**

KS is indebted to the National Institute for Materials Science, Japan Prize Foundation, and Program for Development of Environmental Technology using Nanotechnology of the Ministry of Education, Culture, Sports, Science and Technology (MEXT) for supports. This author acknowledges fruitful discussions with Prof. O. Sugino (The University of Tokyo, Japan), Prof. T. Ishimoto and Prof. M. Tachikawa (Yokohama City University, Japan), and Dr. M.




Melander (University of Jyväskylä, Finland). This work was partially supported by JSPS KAKENHI grant numbers 17K14546 and 19K15527, and TIA collaborative research program [Kakehashi] grant number TK19-002.

**Appendix 1. Fundamental Meanings of Key Electrochemical Parameters: Symmetry Factor, Transfer Coefficient, and Stoichiometric Number**

Symmetry Factor:

Symmetry factor ($\beta$) is related to the gradients of a potential energy surface for the representative points of a reactant and product. $\beta$ concerns only a single charge transfer step. It is usually close to 0.5. Details of $\beta$ depend on a selection of the theoretical framework. However, the core meaning of $\beta$ is regarded as a coefficient to control the transfer of electrical to chemical energies.[66, 120, 121] Starting from this point, for example, Hush saw $\beta$ as a fractional charge on a reacting ion in a transition state.[20] Marcus regarded this coefficient as a potential-dependent term dependent on reorganizations of solvents.[21]

Transfer Coefficient:

In contrast to $\beta$, which is in a pure single charge transfer process, a transfer coefficient ($\alpha$) reflects a result of a series of electrode processes. Parsons analyzed $\alpha$ instead of $\beta$ theoretically to analyze



a RDS in multistep electrochemical reactions.[61] The unified view of $\alpha$ is still under debate, especially on the possibility of more than one electron transfer at a single RDS (e.g. $r \geq 2$). However, we use Eq. 3 with $r = 1$ because one electron transfer is the highest probability from the point of a wide spectrum of theories.

Stoichiometric Number:

A stoichiometric number ($v$) is an indicator to know how many times the RDS takes place for the overall reaction to occur once.[66] It was introduced by Horiuti and Ikushima in 1939.[122]

## Appendix 2. Derivation of Key Equations

Derivation of Eq. 4:

$$K^{\text{H/D}}_{\text{HER}}(\eta) = \frac{k^{\text{H}}(\eta)}{k^{\text{D}}(\eta)} = \frac{j_0^{\text{H}}}{j_0^{\text{D}}} \times \frac{[\text{D}^+]}{[\text{H}^+]} \times \exp\left\{\frac{(\alpha^{\text{D}} - \alpha^{\text{H}})F\eta}{RT}\right\}$$

$\because \quad j_0 = nF[\text{X}]k_0 \quad (\text{X} = \text{H}^+ \text{ or D}^+), \ j_{k,\text{HER}} = nF[\text{H}^+]k_0^{\text{H}}\exp\left\{\frac{-\alpha^{\text{H}}F\eta}{RT}\right\},$

and $j_{k,\text{DER}} = nF[\text{D}^+]k_0^{\text{D}}\exp\left\{\frac{-\alpha^{\text{D}}F\eta}{RT}\right\}.$

Derivation of Eq. 5:

$$K^{\text{H/D}}_{\text{ORR}}(\eta) = \frac{k^{\text{H}}(\eta)}{k^{\text{D}}(\eta)} = \frac{j_0^{\text{H}}}{j_0^{\text{D}}} \times \frac{c_0^{\text{D}}}{c_0^{\text{H}}} \times \exp\left\{\frac{(\alpha^{\text{D}} - \alpha^{\text{H}})F\eta}{RT}\right\}$$



$$\therefore \; j_{0,\text{ORR}}^{\text{H}} \; = \; nk_0^{\text{H}}C_0^{\text{H}}\exp\left\{\frac{-\alpha^{\text{H}}F\eta}{RT}\right\} \; \text{and} \; j_{0,\text{ORR}}^{\text{D}} \; = \; nk_0^{\text{D}}C_0^{\text{D}}\exp\left\{\frac{-\alpha^{\text{D}}F\eta}{RT}\right\}.$$